# IMPACT OF THE EARTH'S OSCILLATIONS ON THE EARTHQUAKES

A. Guglielmi[1], O. Zotov[2]

1 - Schmidt Institute of Physics of the Earth, Russian Academy of Sciences, Moscow, Russia

e-mail: guglielmi@mail.ru

2 - Borok Geophysical Observatory, Schmidt Institute of Physics of the Earth, Russian Academy of Sciences, Borok, Yaroslavl oblast, Russia

e-mail: ozotov@inbox.ru

**Abstract**—In this work we have discovered that the global seismic activity undergoes a modulation with the period of 54 min. We assume that the modulation is caused by the fundamental spheroidal oscillations of the Earth $_0S_2$. The hidden 54-min periodicity of earthquakes was revealed by analyzing of earthquake catalogs using the method of epoch superposition and the spectral method. Initially we discovered some signs of the impact of $_0S_2$ oscillations on the earthquakes accidentally when searching for an antipodal effect of the strong earthquakes. Subsequently we found independent evidence in favor of our hypothesis when analyzing human impact on the seismic activity. Namely, along with the spectral peak of the seismicity at the frequency of 0.277 mHz (which is clearly of anthropogenic origin) we found a more powerful peak at the frequency of 0.309 mHz, i.e. on the period of 54 min, which coincides with the period of the fundamental mode of eigen spheroidal oscillations of the Earth.

# 1. Introduction

It is well known that the earthquakes excite the free oscillations of the Earth as a whole at the resonant frequencies of toroidal and spheroidal eigen oscillations. This fundamental problem is investigated thoroughly both theoretically and experimentally (e.g., see the well-known books [Bullen, 1975; Aki, Richards, 1980; Zharkov, 1986]). We would like to present the arguments in favor of the idea that a reverse process also takes place. Namely, the Earth's free oscillations induce an earthquake activity. We will focus on the spheroidal oscillations $_0S_2$ whose period is 54 min, and we will present evidence that the seismic activity is modulated with this period.

Generally speaking, the idea of induced seismicity is not a new one. Various aspects of the physics of the induced seismicity are widely discussed in the literature (e.g., see [Nikolaev, Vereschagina, 1991; Hill et al., 1993, 2003; Adushkin, Turuntaev, 2005]). The novelty of our result is that we have found a global effect of induced seismicity at the frequency of fundamental oscillations of the Earth. This result was obtained by the analysis of earthquake catalogs using the method of epoch superposition, as well the spectral method.

Initially we discovered some signs of the impact of $_0S_2$ oscillations on the earthquakes accidentally, when searching for an antipodal effect of the strong earthquakes. These signs were found in the vicinities of antiepicenter and epicenter. The corresponding results are presented below in Sections 2 and 3 respectively.

In Section 4 we present an additional proof which was also found to some degree by chance. Namely, we have studied so called Big Ben effect. It manifests itself in the form of hidden strictly periodical human impact on the seismic activity. The effect was initially found by the analysis of vast volumes of data on the earthquakes using the synchronous detection method [Zotov, Guglielmi, 2010; Guglielmi, Zotov, 2012]. After that we tried to reproduce the Big Ben effect by analysis of earthquake catalogs using the spectral method, but faced with an extremely interesting and seemingly strange phenomenon. Along with a spectral peak at the frequency of 0.277 mHz, which is clearly of anthropogenic origin (Big Ben effect), we found a more powerful peak at the frequency of 0.309 mHz, i.e. on the period of 54 min, which coincides with the period of the fundamental mode of eigen spheroidal oscillations of the Earth.

# 2. Antipodal effect

The idea of antipodes was known since antiquity, but the antipodal effect as a wave phenomenon became widely known only in 1957 after launching Sputnik by the Soviet Union. Millions of radio amateurs were receiving and listening the satellite signal. The amplification of the signal during the passage of satellite the neighborhood of antipodal point of the trajectory caused a special interest and widespread enthusiasm [Alpert, 1961]. It is clear that the propagation of surface seismic waves excited by an earthquake leads to a similar effect. Indeed, suppose that a homogeneous elastic sphere is excited by a point source located at the surface the sphere at the point $\theta = 0$. Solutions of the equation for Rayleigh waves are expressed in terms of Legendre functions $P_\nu^\mu(\cos\theta)$. Asymptotic representation of the Legendre functions contains a factor $\sqrt{2/\pi \sin\theta}$ [Gradshteyn, Ryzhik, 1965]. This shows that the amplitude of elastic vibrations increases as we approach the antipodal point $\theta = \pi$ (the antipodal point is sometimes called the antiepicenter). The wave propagates at half the distance round the world for 90 minutes at a characteristic velocity of 3.7 km / s. Thus, increasing the amplitude of oscillations at the antipodal point and its surroundings is expected in about hour and a half after the earthquake.

Because of the unpredictability of earthquakes it is difficult to directly observe the antipodal effect in an individual event. It is therefore reasonable to attempt to find the indirect signs of amplification of surface waves in the antipodal zones by statistical analysis of long series of observations. The necessary initial information we found in the earthquakes catalog of the International Seismological Centre ISC (http://www.isc.ac.uk) and in the earthquakes catalog of the National Earthquake Information Center U.S. Geological Survey USGS / NEIC (http:// neic.usgs.gov / neis / epic / epic_global.html). Of course, using the catalogs as a source of information, we can not directly see the linear antipode effect, i.e. amplification of elastic vibrations with the approach to antipodal point. Analyzing the catalogs we expect to detect the non-linear effect in the form elevated seismic activity in a neighborhood of antipodal points. In other words, our expectations are associated with the concept of induced seismicity (see for example [Nikolaev, Vereschagina, 1991; Hill et al., 1993, 2003; Adushkin, Turuntaev, 2005]). Looking ahead, we say that we were able to detect some signs of increasing the number of weak shocks in the antipodal zone after approximately in 1.5 h after the start of a major earthquake. This result is the indirect indication of a modulation of the seismic activity caused by the spheroidal oscillations of the Earth.

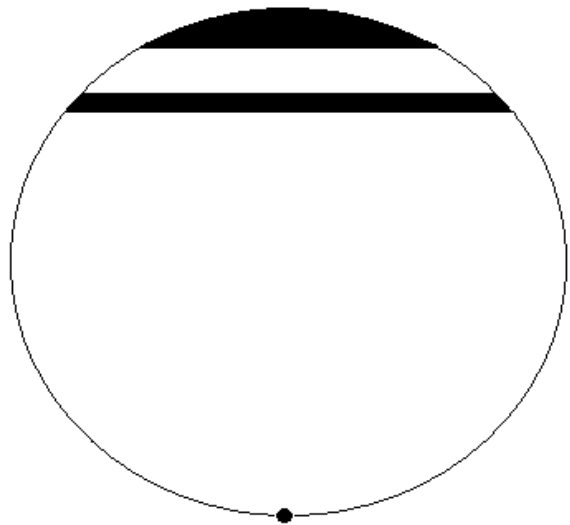

Fig. 1. The earthquake epicenter, the antipodal zone, and a spherical belt (see the text).

Let us choose antipodal zone in the form of a spherical segment of radius $\pi - \theta_0$ centered at the antipodal point (Fig. 1). Here $\theta_0$ is the distance from the epicenter to the boundary of segment. For comparison, we also need the spherical zone, whose area $2\pi R_E^2 (\cos\theta_2 - \cos\theta_1)$ is equal to the area $2\pi R_E^2 (1 + \cos\theta_0)$ of antipodal zones. Here $\theta_{1(2)}$ is the epicentral distance to the distant (nearest) boundary of a spherical zone, and $R_E$ is the radius of the Earth. For given values of $\theta_0$ and $\theta_1$ we find $\theta_2$ from the condition of equality of the areas:

$$\theta_2 = \arccos(1 + \cos\theta_0 + \cos\theta_1). \tag{1}$$

The radius of the spherical segment $\pi - \theta_0$ is selected from the following considerations. On the one hand, the asymptotic relation $P_\nu^\mu(\theta) \propto \sqrt{2/\pi \sin\theta}$ dictates the choice of sufficiently small radius. (Of course, taking into account the fact that this relation does not take into account the diffraction and practically inevitable spherical aberration.) On the other hand, it is desirable that in the antipodal zone was as much as possible earthquakes. We would like to explain why these two conditions are not easily satisfied.

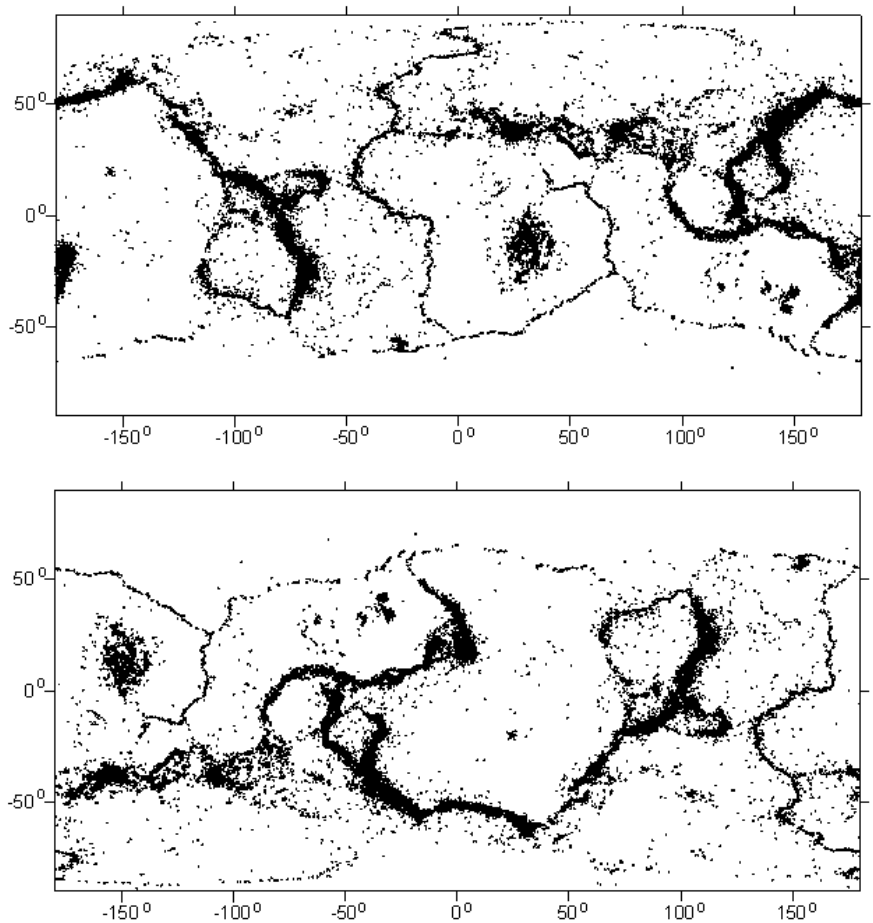

Fig. 2. World map showing the distribution of earthquake epicenters (upper panel), and the distribution of antiepicenters (lower panel).

Fig. 2 shows the two maps of the world. One shows the distribution of earthquake epicenters, and the other shows the distribution of the antipodal points. Let's mentally impose one card on another. It is now clear that the regions of densely located points on one map, generally fall in the regions of sparsely located points on the other card. Therefore, for finding of the induced seismicity in the antipodal zone, it is necessary to choose a sufficiently large radius of the zone. After a series of trials and errors, we chose $\theta_0 = 160°$ and $\theta_1 = 135°$, which correspond to the radius of the antipodal zones equals to $20°$, and the distance from boundary of spherical belt to the antipodal point equals to $45°$. Then the formula (1) gives $\theta_2 = 130.3°$. In other words, the width of the spherical belt equals to $4.7°$, i.e. it is several times smaller than the radius of the antipodal zones (see Fig. 1).

To search for an antipodal effect we used the data on earthquakes specified in the catalog ISC from 1964 to 2006. First of all, 6934 strong earthquakes with magnitudes $M \geq 6$ have been selected from the catalog. The number of such earthquakes is equal to 6934. These earthquakes were used as benchmarks. For the rest of the earthquakes, which were mostly quite weak ($M \leq 1$), the following procedure of selection was applied. The antipodal zone and the spherical belt with the parameters listed above were constructed for each benchmark. Then, from the set of small earthquakes, only those were selected that occurred in the antipodal zone or in the spherical zones within of $\pm 10$ h with respect to the time of corresponding benchmark.

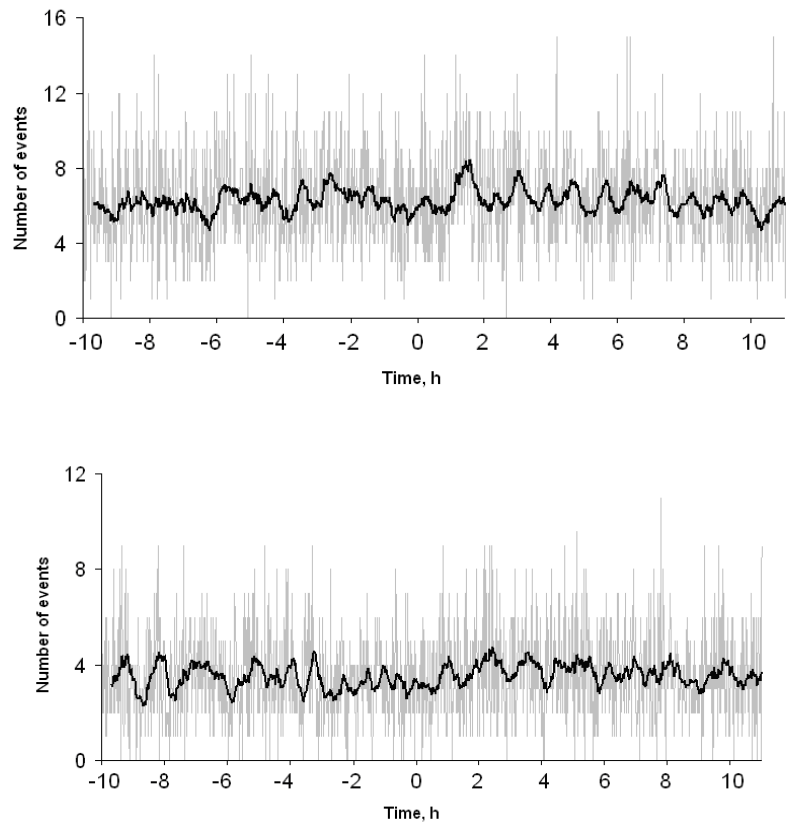

Fig. 3. Distribution of small earthquakes ($M \leq 1$) with respect to the moments of strong earthquakes ($M \geq 6$). The top panel shows the distribution in antipodal zone, and the bottom one

shows the distribution in the spherical belt, which area equals to the area of antipodal zone. Thick curves are 20-points smooth of the initial data.

Fig. 3 shows the result of selection and summation by the method of superposed epoch. Let us focus on the thick curves obtained by smoothing the initial distributions (thin lines) to 20 points. The bottom panel shows the distribution in the spherical belts (4975 events). How it was expected, we find nothing noteworthy here, that would be somehow associated with the zero point. In contrast, on the top panel, which presents the distribution in antipodal zone (8975 events), we see a noticeable maximum at 90 min after zero point. In this regard, we would like to say about a "divine surprise to see the phenomenon predicted by theory, where it was predicted, and as such, as predicted", just as said Anatole Abragam in a different context [Abragam, 1989]. But we are quite aware that a rather weak effect obtained by the method described above, of course, requires independent confirmation. We shall return to this issue in the next section.

Concluding this section we turn our attention at the strange quasi-periodical oscillations of seismicity with period of about one hour, which are especially noticeable on the top panel of Fig. 3. Oscillations have started with a short delay after a strong earthquake and continued for several hours. We discovered these oscillations by accident. They are not directly related to the nonlinear antipodal effect. Nevertheless, below we discuss the reality of oscillations of seismicity, because the period of about one hour is close to the fundamental period of free oscillations of the Earth.

### 3. Aftershocks

The epicenter of the earthquake and antiepicenter, being respectively the nearest and farthest points on the surface of the Earth from the hypocenter, are mutually antipodal points. Propagating from the epicenter with a characteristic velocity 3.7 km / s, the surface elastic waves are returned to the neighborhood of the epicenter in 3 h after the earthquake having made a complete revolution around the Earth. The linear antipodal effect is in essence that the amplitude of elastic vibrations is enhanced in the vicinities of mutually antipodal points. Formally, this is reflected in the fact that the asymptotic of the Legendre functions $P_\nu^\mu(\cos\theta)$ has a pole not only for $\theta = \pi$ (antiepicenter), but also at the $\theta = 2\pi$ (epicenter).

We would like to explain why it is better to seek the nonlinear antipodal effect in the neighborhood of the epicenter than in the vicinity of antiepicenter. From general considerations it is reasonable to assume that the effect is manifested primarily in the areas where the stress-strain state of the Earth's crust is close to the threshold beyond which the mainline rupture of the rocks occurs, and an earthquake starts even under the influence of rather small vibrations. Due to unpredictability of the earthquakes it is difficult to find places in advance. But there is one exception. Paradoxically enough, but we mean the region of aftershocks in the vicinity of a strong earthquake. Indeed, the very fact of a strong earthquake indicates the transition through the critical threshold, and the subsequent aftershocks testify that the accumulated stress of rocks does not completely removed due to the main shock.

Based on these considerations, we tried to detect the antipodal effect in the vicinity of the epicenter in the form of increased activity of aftershocks in about 3 hours after the main shock. For this purpose we used the USGS catalog from which we selected the strong earthquakes with magnitudes $M \geq 7$, and took into account all foreshocks and aftershocks with magnitudes $6 \leq M < 7$ in the epicentral zones of $10^o$ radius, and in the time interval $\pm 10$ h relative to the moment of the main shock. We accumulated 500 main shocks and 234 foreshocks and aftershocks for the period from 1973 to 2010. We then used the method of superposed epochs taking as the zero reference point the time of the main shock.

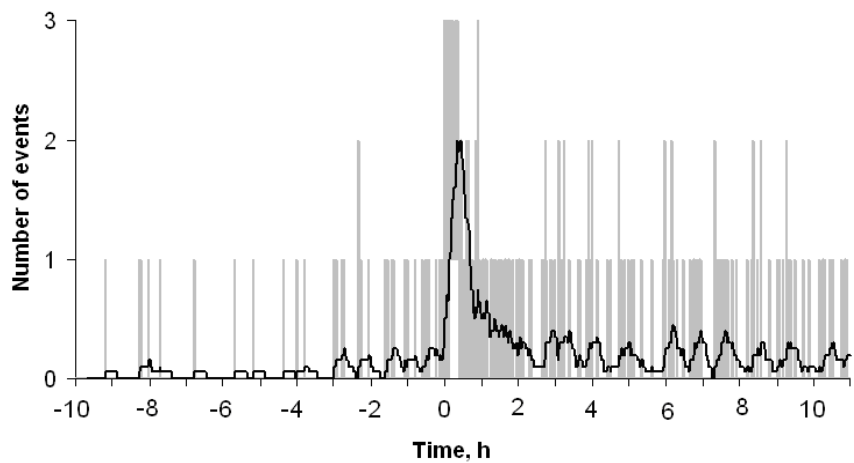

Fig. 4. Dynamics of the foreshocks and aftershocks in the epicentral zones of the strong earthquakes. The thick curve is 20-point smooth of the initial data.

The result of the analysis is shown in Fig. 4. We see that morphologically the evolution of seismicity after the main shock is naturally divided into two phases, namely the phase of spontaneous activity (0 - 2.5 h) and the phase of induced activity (2.5 h - 10 h). On average, the maximum activity of aftershocks is observed during the first hour after the strong earthquake. After that comes some sedation with a minimum of aftershocks at the 2.5 h. Next there is some increasing activity which leads to a "forked" maximum in about 3 h.

Disregarding details, it seems that our expectations have justified to some extent. At the same time we found something quite unexpected. Namely, Fig. 4 shows quite clearly the oscillations of the aftershock activity with a quasi-period of about 55 minutes which is very close to the period of natural oscillations of the Earth $_0S_2$. This gives grounds to suggest that the oscillations of the Earth as a whole leads to a nonlinear modulation of the seismicity.

## 4. Spectrum of the global seismicity

Now therefore, the studies of antipodal effects in the vicinities of epicenter and antiepicenter had led us to an unexpected result. Namely, we found signs of the modulation of seismic activity under the influence of spheroidal oscillations of the Earth. In this section we show that this modulation is global in a certain sense.

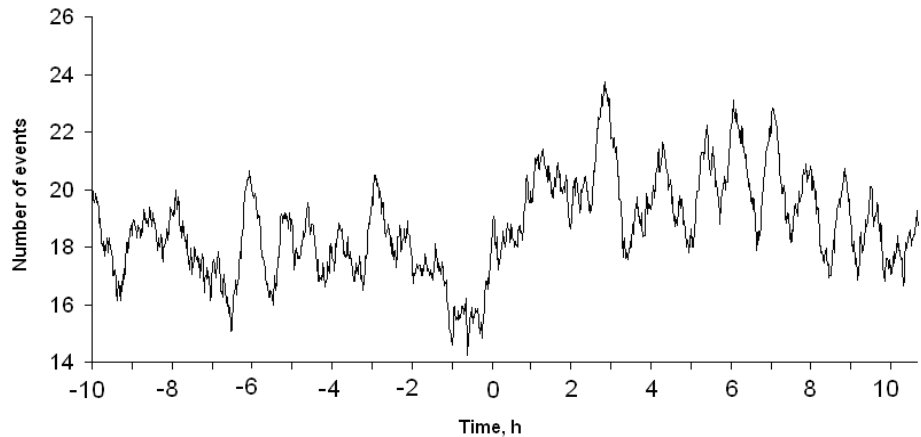

Fig. 5. The distribution of earthquakes in time, similar to that shown on the top panel of Fig. 3. The radii of the antipodal zones are taken here equal to $30^o$.

We first look at Fig. 5. It shows the distribution of small earthquakes in time relative to strength earthquakes in the same way as it was done in the upper panel of Fig. 3. The difference is that we have increased the radius of the antipodal zones from $20^o$ to $30^o$. Increase of the search area has led us to a noticeable increase in the number of small earthquakes falling into the antipodal zones (23 607 events instead of 8 975). As a result, we are seeing a train of damped oscillations more clearly than in Fig. 3. The oscillation period is about 55 minutes according to an indicative assessment. This observation gave us the idea to expand the area of search as much as possible. We rejected the introduction of reference points, and we have removed all the spatial and temporal constraints in the selection of events from the earthquake catalog.

For the analysis it has been used all earthquakes with magnitudes $M \geq 1$ which were registered in the USGS catalog from 1973 to 2010 at any point on the Earth's surface. The total number of earthquakes that have occurred over 38 years and were listed in the catalog equals to 536 000. Let us perform a spectral analysis of the sequence of these earthquakes. To this purpose, we assign zero for each second interval in the course of the 38 years if nowhere at this interval where was no earthquakes, or a positive integer $\nu_j$, if somewhere at this interval where was $\nu_j$ of earthquakes. We represent the dynamics of earthquakes as a series

$$n(t) = \sum_{j=1}^{N} \nu_j \delta(t - t_j), \qquad (2)$$

in which $t_j$ is the beginning of second interval, which is assigned a number $t_j$, $N$ is the total number of such intervals, and $\delta(t)$ is the Dirac delta function. Let us represent the function $n(t)$ as a Fourier integral:

$$n(t) = \int_{-\infty}^{\infty} n_\omega \exp(-i\omega t) \frac{d\omega}{2\pi}. \qquad (3)$$

Here the spectral component $n_\omega$ is given by the expression

$$n_\omega = \int_{-\infty}^{\infty} n(t) \exp(i\omega t) dt. \qquad (4)$$

Substituting (2) into (4), we find

$$n_\omega = \sum_{j=1}^{N} \nu_j \exp(i\omega t_j). \qquad (5)$$

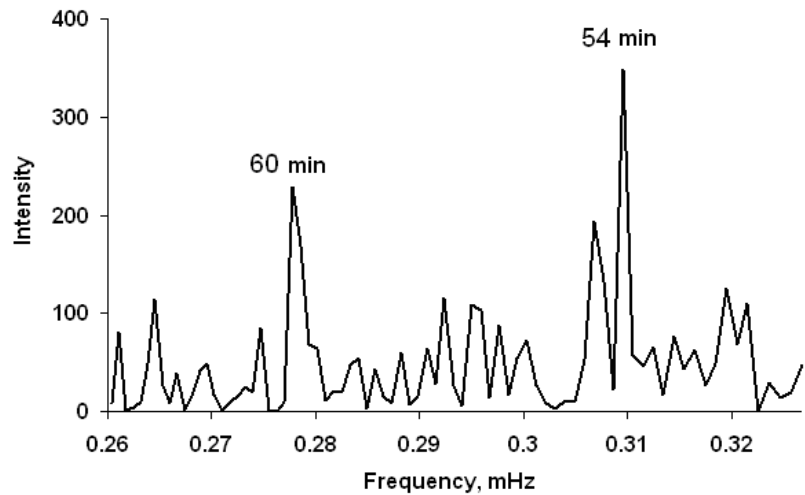

Fig. 6. Spectrum of global seismic activity from 1973 to 2010.

Fig. 6 shows the intensity of the Fourier components $|n_\omega|^2$. We see the spectral peak at the frequency of 0.277 mHz. It corresponds to the period of 60 min. We believe that this peak is of anthropogenic origin. In other words, the peak at the period of 60 min relates somehow with the impact the technosphere on the lithosphere. In this regard, we note that Fig. 6 shows only one fragment of spectrum that we are interested in this work. The full spectrum contains also the spectral peaks at the periods of 15 min, 24 h and 7 d which are of "human" origin (see the papers [Zotov, Guglielmi, 2010; Guglielmi, Zotov, 2012] for more details concerning a human impact on the geospheres).

We see in Fig. 6 that the spectrum of seismic activity along with a peak at the frequency of 0.277 mHz, which is clearly of anthropogenic origin, there is a strong peak at the frequency of 0.309 mHz. The frequency of 0.309 mHz corresponds to the period of 54 min, which coincides with the period of the Earth's fundamental oscillations $_0S_2$. This is an indication that the Earth's free oscillations modulate the global seismic activity.

## 5. Conclusion

In this work we have discovered that the global seismic activity undergoes a modulation with the period of 54 min. We assume that the modulation is caused by the fundamental spheroidal oscillations of the Earth $_0S_2$. The specific physical mechanisms responsible for the interaction between the Earth's spheroidal oscillations and the earthquakes, yet to be clarified in the further studies.

To reveal the hidden 54-min periodicity of earthquakes we used the data for a very long period of observations (several decades). One of the tasks of further research is to develop the methods to study the impact of the Earth's oscillations on the seismicity by using observations over relatively short intervals. The solution to this problem would open interesting perspectives for the monitoring of the Earth's crust as a whole according to the observations of the modulation of induced seismicity and data on the free oscillations of the Earth over the same period of time.

In conclusion let us touch the question of whether it is possible to use the described property of earthquakes in astroseismology, particularly in the seismology of pulsars. One would think, why not try to identify the resonance peaks in the sequence of pulsar quakes and, thus, estimate the periods of elastic vibrations of a pulsar? However, to our knowledge, the corresponding series of observations of sufficient length are absent at present.

**Acknowledgments**

We thank the staff of ISC and USGS/NEIC for providing the catalogues of earthquakes. We are grateful to B.I. Klain, A.S. Potapov, L.E. Sobisevich, and A.L. Sobisevich for their interest to our work and valuable comments. The work was partially supported by the Program № 4 of the Presidium of RAS.

**Figure Captions**

Fig. 1. The earthquake epicenter, the antipodal zone, and a spherical belt (see the text).

Fig. 2. World map showing the distribution of earthquake epicenters (upper panel), and the distribution of antiepicenters (lower panel).

Fig. 3. Distribution of small earthquakes ($M \leq 1$) with respect to the moments of strong earthquakes ($M \geq 6$). The top panel shows the distribution in antipodal zone, and the bottom one shows the distribution in the spherical belt, area of which is equal to the area of antipodal zone. Thick curves are 20 points smooth of the initial data.

Fig. 4. Dynamics of the foreshocks and aftershocks in the epicentral zones of the strong earthquakes. The thick curve is 20 points smooth of the initial data.

Fig. 5. The distribution of earthquakes in time, similar to that shown on the top panel of Fig. 3. The radii of the antipodal zones are taken here equal to $30^o$.

Fig. 6. Spectrum of global seismic activity from 1973 to 2010.